\documentclass[12pt]{article}
\usepackage{graphicx}

\newcommand\pubdate{\today}

\newcommand\pubnumber{}

\def\Title#1{\begin{center} {\Large #1 } \end{center}}
\def\Author#1{\begin{center}{ \sc #1} \end{center}}
\def\Address#1{\begin{center}{ \it #1} \end{center}}

\newcommand\pubblock{\rightline{\begin{tabular}{l} \pubnumber\\
         \pubdate  \end{tabular}}}
\newenvironment{Abstract}{\begin{center}{\bf Abstract}\end{center} \bigskip \begin{quotation}  }{\end{quotation}}
\newenvironment{Presented}{\begin{quotation} \begin{center} 
             PRESENTED AT\end{center}\bigskip 
      \begin{center}\begin{large}}{\end{large}\end{center} \end{quotation}}

\def\beq{\begin{equation}}
\def\eeq#1{\label{#1}\end{equation}}
\def\eeqn{\end{equation}}

\def\beqa{\begin{eqnarray}}
\def\eeqa#1{\label{#1}\end{eqnarray}}
\def\eeqan{\end{eqnarray}}

\let\bar=\overbar

\def\Dslash{\not{\hbox{\kern-4pt $D$}}}
\def\dslash{\not{\hbox{\kern-2pt $\del$}}}

\def\msb{{\bar{\ssstyle M \kern -1pt S}}}

\begin{document}
\begin{titlepage}
\pubblock

\vfill

%%%%%%%%%%%%%%%%%%%%%%%%%%%%%%%%%%%%%%%%%%%%%%%%%%%%%%%
%%MODIFY
%%%%%%% TITLE, AUTHOR, ADDRESS 
%%%%%%%%%%%%%%%%%%%%%%%%%%%%%%%%%%%%%%%%%%%%%%%%%%%%%%%

\Title{New physics limits from kaon decays}
\vfill
\Author{G. Ruggiero}  
\Address{CERN, Geneva, Switzerland}
\vfill

%%%%%%%%%%%%%%%%%%%%%%%%%%%%%%%%%%%%%%%%%%%%%%%%%%%%%%%
%%MODIFY
%%%%%%% Abstract
%%%%%%%%%%%%%%%%%%%%%%%%%%%%%%%%%%%%%%%%%%%%%%%%%%%%%%%

\begin{Abstract}
Searches for lepton flavour violation and lepton number violation in kaon decays by the NA62 and NA48/2 experiments
at CERN are presented. A new measurement of the ratio of charged kaon leptonic decay rates $R_K=\Gamma(K_{e2})/\Gamma(K_{\mu2})$
to sub-percent relative precision is discussed. An improved upper limit on the lepton number violating $K^\pm\rightarrow\pi^\mp\mu^\pm\mu^\pm$
decay rate is also reported. The future 10\% precision measurement of the branching ratio of the ultra-rare kaon decay $K^+\rightarrow\pi^+\nu\bar{\nu}$
with the NA62 experiment is finally reviewed.
\end{Abstract}

\vfill

\begin{Presented}
The Ninth International Conference on\\
Flavor Physics and CP Violation\\
(FPCP 2011)\\
Maale Hachamisha, Israel,  May 23--27, 2011
\end{Presented}
\vfill

\end{titlepage}
\def\thefootnote{\fnsymbol{footnote}}
\setcounter{footnote}{0}
%

%%%%%%%%%%%%%%%%%%%%%%%%%%%%%%%%%%%%%%%%%%%%%%%%%%%%%%%
%%%%%%% Article body
%%%%%%%%%%%%%%%%%%%%%%%%%%%%%%%%%%%%%%%%%%%%%%%%%%%%%%%

\section{Introduction}
In the Standard Model (SM) the decays of charged pseudo scalar mesons ($P$) into lepton neutrino are helicity suppressed. 

Supersymmetric new physics models, like certain 2-Higgs doublet models (e.g. 2HDM type II) \cite{Deschamps:2009rh}, predict sizable deviations from the SM via new physics 
contributions already at tree level. In these frameworks the supersymmetric parameters $\tan\beta$ (the
ratio of the two Higgs vacuum expectation values) and $M_{H^+}$ (the mass of the charged Higgs) usually describe the new physics contributions. 
The dependence of these decay rates on $\tan^2\beta$ and $(M_P/M_{H^+})^2$ naturally enhances the sensitivity of the $B$ mesons to new physics, like the 
$B^+\rightarrow\tau^+\nu_\tau$ decay. Although similar new physics contributions for $K^+\rightarrow l^+\nu_l$ could result in 100 times smaller effects, 
leptonic kaon decays still offer the opportunity to search for new physics thanks to a very high experimental precision \cite{Antonelli:2010yf}. However, precision
 measurements in this sectors clash with the poor knowledge of the hadronic matrix elements which severely limits the theoretical prediction of 
$\Gamma(P^+\rightarrow l^+\nu_l)$. This uncertainty largely cancels in the ratio of the rates of these decays into different lepton families 
(e.g. with $l=e,\mu$), like the parameter $R_K=\Gamma(K_{e2})/\Gamma(K_{\mu2})$. Processes with $e\nu$ or $\mu\nu$ in the final state, on the other side, 
are experimentally accessible only in the $\pi$ and $K$ sector. 

The SM prediction for $R_K$ inclusive of internal bremsstrahlung (IB) radiation is \cite{Cirigliano:2007xi}:
\begin{equation}
R_K^{SM}\,=\,\left(\frac{M_e}{M_\mu}\right)^2\left(\frac{M^2_K-M^2_e}{M^2_K-M^2_\mu}\right)^2(1+\delta R_{QED})\,=\,(2.477\pm0.001)\times10^{-2}, 
\end{equation}
where $\delta R_{QED}$ is an electromagnetic correction due to the IB and structure-dependent effects.
Deviations of $R_K$ from the SM require new physics models with sources of lepton flavour violation (LFV) \cite{Masiero:2005wr,Masiero:2008cb}. 
Within the MSSM, for example, the LFV sources appear at the one loop level via the exchange of the charged Higgs boson coupled with a right-handed 
slepton loop. The dominant contribution is
\begin{equation}
R_K^{LFV}\,\simeq\,R_K^{SM}\left[1+\left(\frac{M_K}{M_H}\right)^4\left(\frac{M_\tau}{M_e}\right)^2|\Delta^{31}_R|^2\tan^6{\beta}\right],  
\end{equation}
where $|\Delta_R^{31}|$ is the mixing parameter between the superpartners of the right-handed leptons, which can reach values up to $10^{-3}$. After an appropriate tuning 
of the new physics parameters, the effect on $R_K$ could be up to \% level without contradicting any experimental constraints. Already in the 1970s several experiments
measured $R_K$ \cite{Clark:1972nq,Heard:1974kk,Heintze:1976qf}, while the present PDG value \cite{Nakamura:2010zzi}, $R_K=(2.493\pm0.031)\times10^{-5}$, is largely 
dominated by a recent result from KLOE \cite{Ambrosino:2009rv}. A new measurement of $R_K$ based on a part of a data sample collected by the NA62 experiment 
(phase I) at CERN in 2007 \cite{Lazzeroni:2011} is reported here (section \ref{sec:rk}).

Kaon decays may also contribute to the search for lepton number violation via decays like $K^\pm\rightarrow\pi^\mp\mu^\pm\mu^\pm$. They violate the lepton number by 2 units and
can proceed only if the $\nu$ is a Majorana particle: consequently the limit on their branching ratio provide constraints on the effective Majorana neutrino mass \cite{Zuber:2000vy}.
Such processes were already studied experimentally by the BNL E865 experiment in 1997 \cite{Appel:2000tc}. The NA48/2 experiment at CERN collected a $\pi\mu\mu$ 
sample about 8 times larger than the one from E865. It allows improving the limits on the $K^\pm\rightarrow\pi^\mp\mu^\pm\mu^\pm$ process significantly \cite{Batley:2011zz}
(section \ref{sec:pmm}).

Among the many rare flavour changing neutral current $K$ and $B$ decays, the ultra rare decays $K\rightarrow\pi\nu\bar{\nu}$ 
play a key role in search for new physics through underlying mechanisms of flavour mixing. The SM 
branching ratio can be computed to an exceptionally high degree of precision and the prediction 
for the $K^+\rightarrow\pi^+\nu\bar{\nu}$ channel is $(7.81\pm0.75\pm0.29)\times10^{-11}$ \cite{Brod:2010hi}. The first error 
comes from the uncertainty on the CKM matrix elements, the second one is the pure theoretical uncertainty. 
The extreme theoretical cleanness of these decays remain also in new physics scenarios like Minimal Flavour Violation (MFV) 
\cite{Isidori:2006qy} or non-MFV models \cite{Blanke:2009am} and even not large deviations from the SM value (for example 
around 20\%) can be considered signals of new physics. The decay $K^+\rightarrow\pi^+\nu\bar{\nu}$ has been observed by the 
experiments E787 and E949 at the Brookhaven National Laboratory and the measured branching ratio is $1.73^{+1.15}_{-1.05}\times10^{-10}$
\cite{Artamonov:2009sz}. However only a measurement of the branching ratio with at least 10\% accuracy can be a 
significant test of new physics. This is the main goal of the NA62 experiment at CERN-SPS \cite{Anelli:2005ju} (section \ref{sec:pvv}). 

\section{The NA48/2 and NA62 (phase I) Experiments}
The NA48/2 and NA62 (phase I) experiments at CERN collected data in 2003-04 and 2007-08 using the same beam line and experimental set-up, 
respectively. NA48/2 aimed to the study of the CP violation in the decay of the charged kaons into three pions \cite{Batley:2007yfa}, NA62 to the 
measurement of the above defined $R_K$ ratio. They were fixed target experiments which used a 400 GeV/c primary proton beam, extracted 
from the SPS accelerator at CERN, which produced a secondary charged kaon beam after impinging on a beryllium target. A 100 m long beam line
selected the momentum of the secondary beam to $(60\pm3)$ GeV/c in 2003-04 and $(75\pm2)$ GeV/c in 2007-08.
Finally the beams entered a decay volume, housed in a 100 m long vacuum tank. With a primary beam intensity of about $7\times10^{11}$ protons 
per SPS spill of 4.8 s duration, the positive (negative) beam flux at the entrance of the decay volume was $3.8\times10^7$ ($2.6\times10^7$) particles 
per pulse. The fraction of kaons decaying in the decay volume was about 20\%, depending on the beam energy.

The detector was designed to see the charged and neutral products of the kaons decaying in the vacuum region. A magnetic
spectrometer tracked the charged particles. It was housed in a tank containing He and separated from the vacuum region by a 
Kevlar window. An aluminum beam pipe of 16 cm diameter with vacuum inside, traversed the spectrometer and allowed the not
decayed beam particles passing through without touching the sensitive detector volume. The spectrometer consisted of four 
drift chambers (DCH) separated by a dipole magnet, which gave to the charged particles an horizontal transverse momentum kick 
of 120 MeV/c (265 MeV/c) 2003-04 (2007-08). The momentum resolution was $\sigma_p/p=(1.02\oplus0.044\times p)$\% in 2003-04
and $\sigma_p/p=(0.48\oplus0.009\times p)$\% ($p$ in GeV/c) in 2007-08.

An hodoscope (HOD), made of two orthogonal planes of 64 plastic scintillator slabs each, followed the magnetic spectrometer. It  
provided the time reference for the other detectors and the main trigger for the events with charged particles.

An electromagnetic calorimeter (LKr), placed after the hodoscope, was used for photon detection and particle identification. 
It was a quasi-homogeneous calorimeter with liquid kripton as active material. A system of Cu-Be ribbons electrodes allowed 
the collection of the ionization signal. In total 13248 projective cells segmented the active volume transversely to the beam 
axis. The total length of the detector corresponded to about 27 $X_0$. The measured energy resolution was 
$\sigma(E)/E=0.032/\sqrt(E)\oplus0.09/E\oplus0.0042$ ($E$ in GeV). 
 
An hadronic calorimeter (HAC) and a muon detector (MUV) followed the electromagnetic calorimeter.

A detailed description of the NA48/2 layout can be found elsewhere \cite{Fanti:2007vi}. 

\section{Measurement of $R_K$ with NA48/2}
\label{sec:rk}
The measurement of $R_K$ has been performed using 40\% of the data collected in 2007 by NA62. The analyzed data contained only positive kaons. 

The measurement relied on counting the numbers of reconstructed $K_{e2}$ and $K_{\mu2}$ candidates collected simultaneously. Consequently $R_K$ did not 
depend on the absolute kaon flux and the ratio allowed for a first order cancellation of several systematic effects, like reconstruction and trigger 
efficiencies and time dependent biases. The basic formula is:
\begin{equation}
R_K\,=\,\frac{1}{D}\cdot\frac{N(K_{e2})-N_B(K_{e2})}{N(K_{\mu2})-N_B(K_{\mu2})}\cdot\frac{A(K_{\mu2})f_\mu\epsilon(K_{\mu2})}{A(K_{e2})f_e\epsilon(K_{e2})}\cdot\frac{1}{f_{LKr}}.
\end{equation}
Here $N(K_{l2})$ and $N_B(K_{l2})$ are the number of the selected $K_{l2}$ events and the expected number of background events, respectively; $D$ is the
downscaling factor applied to the $K_{\mu2}$ trigger; $A(K_{l2})$ the geometrical acceptance of the selected $K_{l2}$; $f_e$ and $f_\mu$ the identification 
efficiencies of electrons and muons, respectively; $\epsilon(K_{l2})$ the trigger efficiencies for the selected $K_{l2}$; $f_{LKr}$ the global LKr efficiency. 

A detailed Monte Carlo simulation (MC) was developed, including beam line optics and time-dependent detector inefficiencies. The computation of the acceptance 
correction $A(K_{\mu2})/A(K_{e2})$ and the geometrical part of the acceptance entering in the background computation relied on MC. The particle identification 
efficiencies, the readout and the trigger efficiencies, instead, were measured directly on data.  
Because both the signal acceptance and the background depended strongly on the lepton momentum, the measurement was performed in bins of this observable by dividing
the range between 13 and 65 GeV/c in 10 intervals.

A large part of the selection was in common between $K_{e2}$ and $K_{\mu2}$, because of the similar single-track topology. Exactly one positive track in the final state, 
reconstructed in the spectrometer and whose extrapolation passed through the downstream detector acceptance, was required; it had to have a momentum within 13 and 65 GeV 
and the reconstructed longitudinal position of the kaon decay vertex had to be located within the fiducial decay region. Events with deposits of energy greater than 2 GeV not 
associated with the charged tracks in the LKr calorimeter were rejected in order to further suppress backgrounds with photons in the final state.

The event kinematics and the lepton identification were effective to separate $K_{e2}$ and $K_{\mu2}$. The $M^2_{miss}=(P_{K}-P_{l})^2$ characterized completely 
the kinematics of the single track decays. Here $P_K$ and $P_l$ are the kaon and lepton 4-momenta respectively; the average $P_K$ was measured on spill basis using the 
$K^+\rightarrow\pi^+\pi^+\pi^-$decays; $P_l$ was computed in the electron or muon mass hypothesis. A cut around the $M^2_{miss}$ peak, according to the 
$M^2_{miss}$ resolution and dependent on the lepton momentum, selected the $K_{l2}$ candidates.
The ratio $E/p$ of the track energy deposit in the LKr calorimeter to its momentum measured by the spectrometer, identified
positrons ($0.95<E/p<1.1$) and muons ($E/p<0.85$). 

\begin{figure}[htb]
\centering
\includegraphics[width=0.45\textwidth]{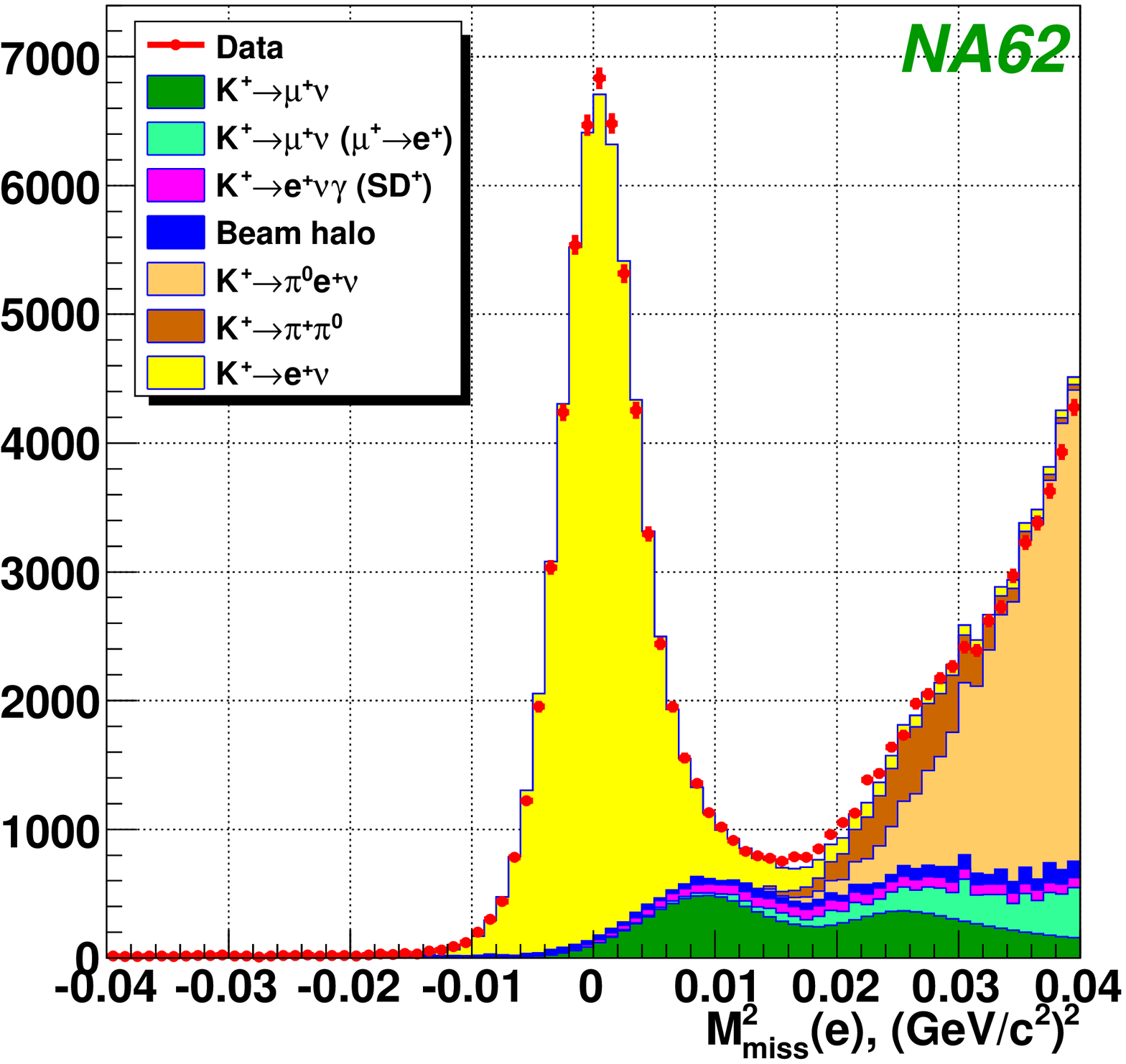}
\includegraphics[width=0.45\textwidth]{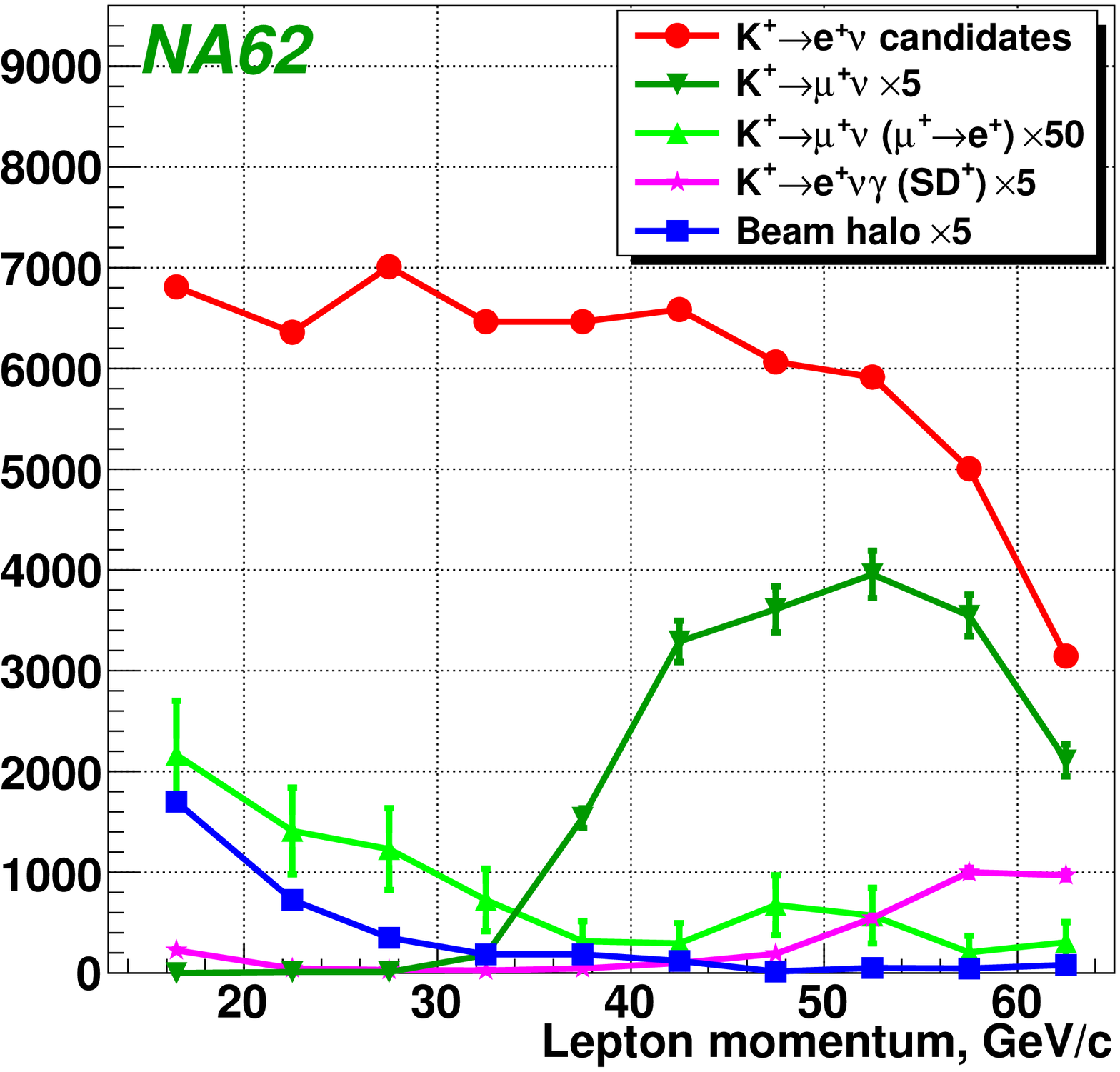}
\caption{Left: reconstructed $M^2_{miss}(K_{e2})$ for $K_{e2}$ and backgrounds. Right: lepton momentum distributions of the $K_{e2}$ candidates and the dominant
backgrounds.}
\label{fig:m2miss}
\end{figure}

The numbers of selected $K_{e2}$ and $K_{\mu2}$ candidates were 59813 and $1.803\times10^7$, respectively. The total background was $8.71\pm0.24$\%. The 
$M^2_{miss}(K_{e2})$ and the background contamination as a function of lepton momentum are shown in figure \ref{fig:m2miss}. 

The background was strongly momentum dependent. In particular for momenta higher than 35-40 GeV/c, the $K_{e2}$ kinematics resembled more and more the $K_{\mu2}$ one 
and $K_{\mu2}$ with a muon mis-identified as a positron became the largest background source. The accuracy of its evaluation was critical to keep the total systematic 
uncertainty smaller than the statistical one. The 'catastrophic' bremsstrahlung in or in front of the LKr was the dominant source of the muon-positron mis-identification. 
The corresponding probability $P_{\mu e}$ was measured on data as a function of lepton momentum. During a first period of the 2007 run, data were taken with a 9.2 $X_0$ 
lead wall in front of the LKr, covering about 20\% of the total geometrical acceptance. This set-up allowed the collection of a muon sample free from the about $10^{-4}$ 
contamination due to $\mu\rightarrow e$ decays. 
The sample used for the measurement of $R_K$, however, was taken without the lead wall. 
Let $P^{Pb}_{\mu e}$ be the probability of muon mis-identification in presence of the lead wall and $P_{\mu e}$ the one without: because of ionization energy loss and 
bremsstrahlung in lead $P_{\mu e}$ and $P^{Pb}_{\mu e}$ differed significantly. Consequently, the value measured with the wall was corrected for using a dedicated simulation 
based on Geant4 \cite{Agostinelli:2002hh}.
The measured $P^{Pb}_{\mu e}$ varied in the range of (3-5)$\times10^{-6}$ according to the muon momentum and was in agreement with the simulation within the uncertainties 
(about 10\% from simulation). The correction $f_{Pb}$ varied from $+10\%$ to $-20\%$ of 
$P^{Pb}_{\mu e}$ depending on lepton momentum. Its uncertainty was around 2\%. The $K_{\mu2}$ background in the $K_{e2}$ sample integrated over lepton momentum 
was $(6.11\pm0.22)\%$. It was computed using the $P^{Pb}_{\mu e}$ measured on data and $f_{Pb}$ from simulation. This combination allowed the minimization of the total 
uncertainty. 

The other background components coming from kaon decays were evaluated using the MC simulation. 
The beam halo background induced by halo muons undergoing decay in flight or mis-identified, was measured by reconstructing positive $K_{e2}$ among data collected from 
$K^-$ beam with the $K^+$ beam blocked while its halo was not. The size of the control sample limited the evaluation of the background uncertainty. 

The acceptance correction was evaluated using MC. The contribution to the $K_{e2}$ acceptance due to the radiative $K^+\rightarrow e^+\nu\gamma$ inner bremsstrahlung
process was taken into account following \cite{Bijnens:1992en,Weinberg:1965nx,Gatti:2005kw}. The bremsstrahlung suffered by the positrons in the material upstream 
of the spectrometer magnet, induced about 6\% loss of $K_{e2}$ acceptance, as a consequence of the $M^2_{miss}$ cut. The effect was computed by studying spectra and 
rates of bremsstrahlung photons produced by 25 GeV/c (40 GeV/c) electron (positron) beam steered into the DCH acceptance, collected by NA48/2 in 2006 (2004). The 
knowledge of the helium purity in the spectrometer tank was the second largest source of systematic uncertainty.

The $R_K$ value was extracted from a $\chi^2$ fit to the measurements in the lepton momentum bins, taking into account the bin-to-bin correlations between the 
systematic uncertainties. Table \ref{tab:results} summarizes the uncertainties. All the assigned systematic errors were checked a posteriori by varying the selection
criteria and the analysis procedure. 
\begin{figure}[htb]
\centering
\includegraphics[width=0.45\textwidth]{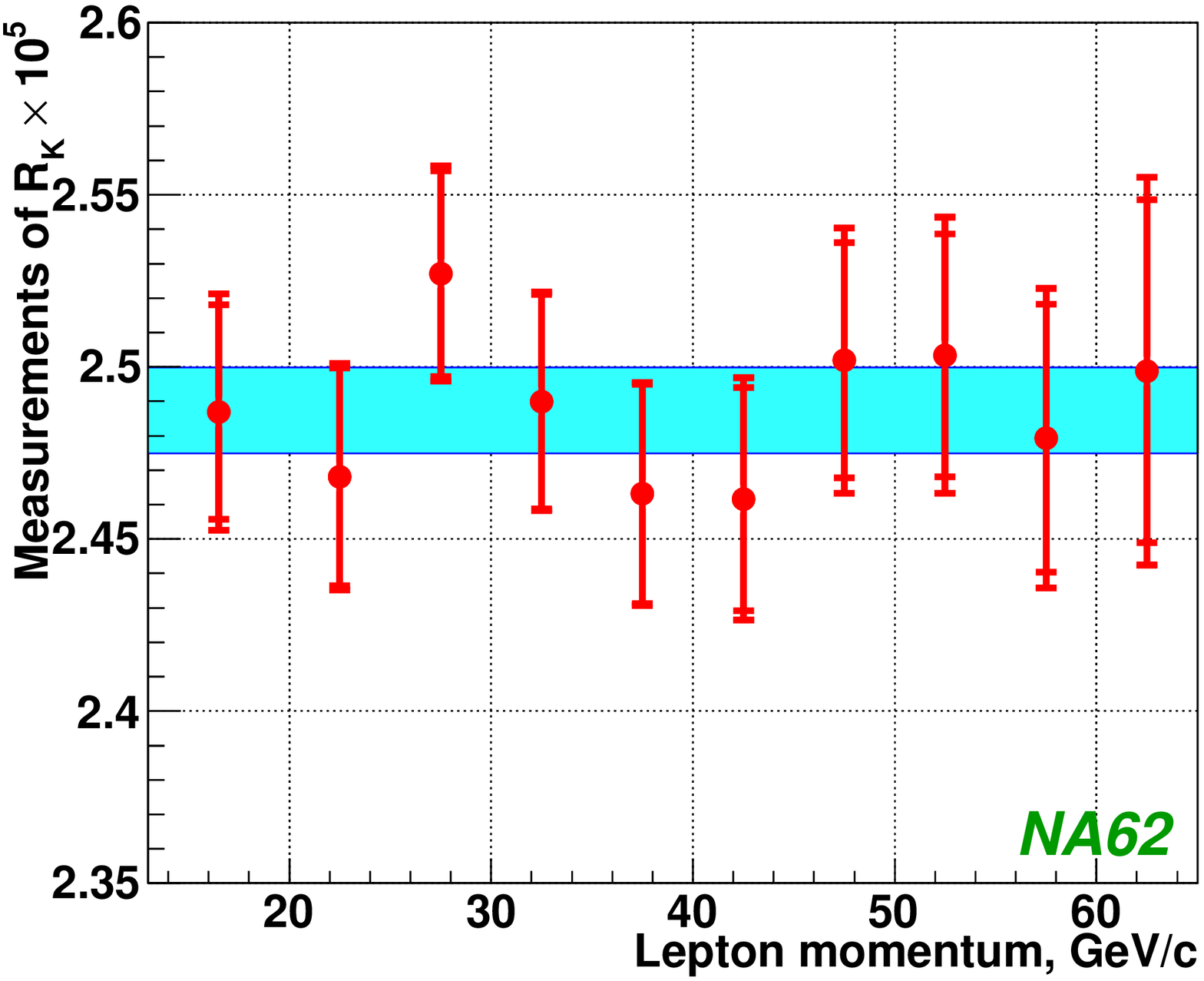}
\includegraphics[width=0.38\textwidth]{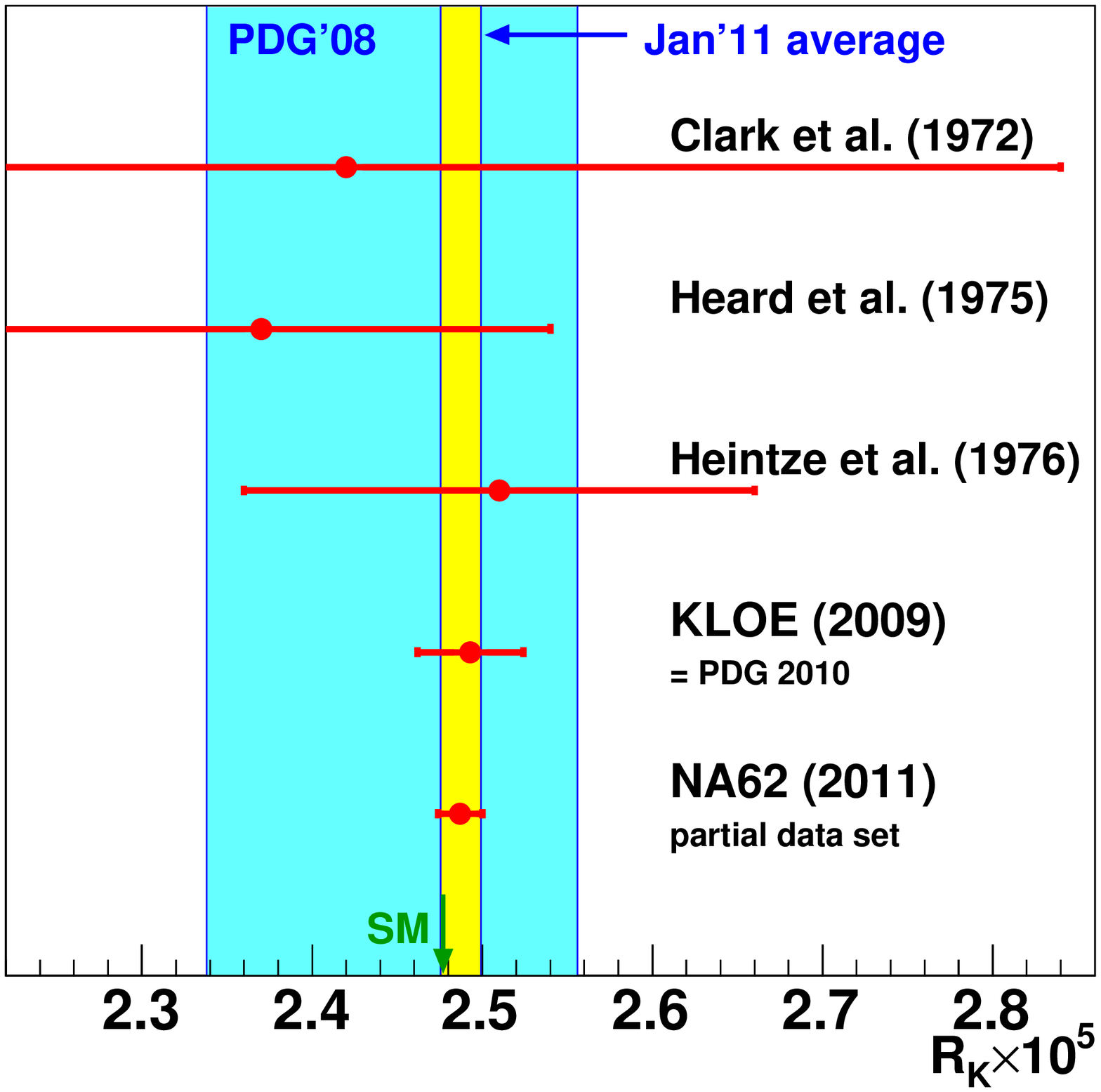}
\caption{Left: measurements of $R_K$ in lepton momentum bins. The band indicates the average $R_K$ and its total uncertainty. Right: the new world average including 
the present result.}
\label{fig:result}
\end{figure}
\begin{table}[!hbtp]
\begin{center}
\begin{tabular}{l|c}  
\hline\hline
 Source                 & $\delta R_K\times10^5$ \\ \hline
 Statistical          & 0.011 \\ \hline
 $K_\mu2$ background    & 0.005 \\
 Other background from $K$ decays & 0.001 \\
 Beam halo background & 0.001 \\
 Helium purity & 0.003 \\
 Acceptance correction & 0.002 \\
 Spectrometer alignment & 0.001 \\
 Positron identification efficiency & 0.001 \\
 1-track trigger efficiency & 0.002 \\
 LKr readout inefficiency & 0.001 \\
 Total systematic & 0.007 \\ \hline
 Total & 0.013 \\
\hline\hline
\end{tabular}
\caption{Summary of the uncertainties on $R_K$.}
\label{tab:results}
\end{center}
\end{table}

The result is ($\chi^2/$ndf = 3.6/9):
\begin{equation}
  R_K\,=\,(2.487\pm0.011_{stat}\pm0.007_{syst})\times10^{-5}\,=\,(2.487\pm0.013)\times10^{-5}.
\end{equation} 
The individual results in lepton momentum bins and the new world average are presented in figure \ref{fig:result}.

\section{Search for lepton number violation with NA48/2}
\label{sec:pmm}
The search for $K^\pm\rightarrow\pi^\mp\mu^\pm\mu^\pm$ decay was performed on the NA48/2 2003-04 data, using the $K^\pm\rightarrow\pi^\pm\pi^+\pi^-$ ($K_{3\pi}$) decay 
as a normalization channel. A three-track event topology was required, with tracks compatible with pion ($E/p<0.95$) or muon hypothesis ($E/p<0.2$). Hits in the muon 
veto in time with the tracks, identified the presence of muons in the final state. The muon identification inefficiency was measured to be below 2\% for momentum greater
than 10 GeV/c. 
\begin{figure}[htb]
\centering
\includegraphics[width=0.45\textwidth]{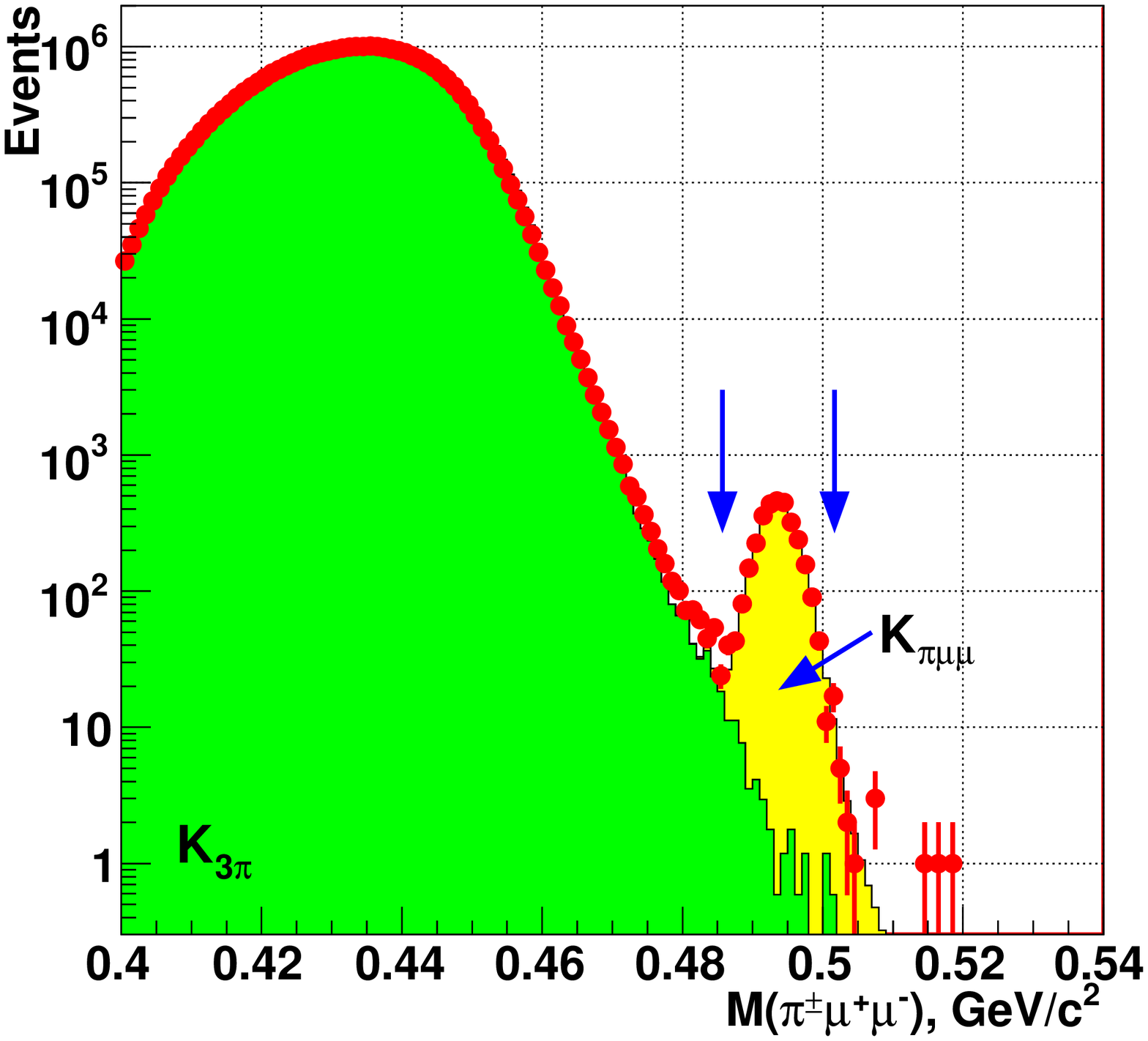}
\includegraphics[width=0.45\textwidth]{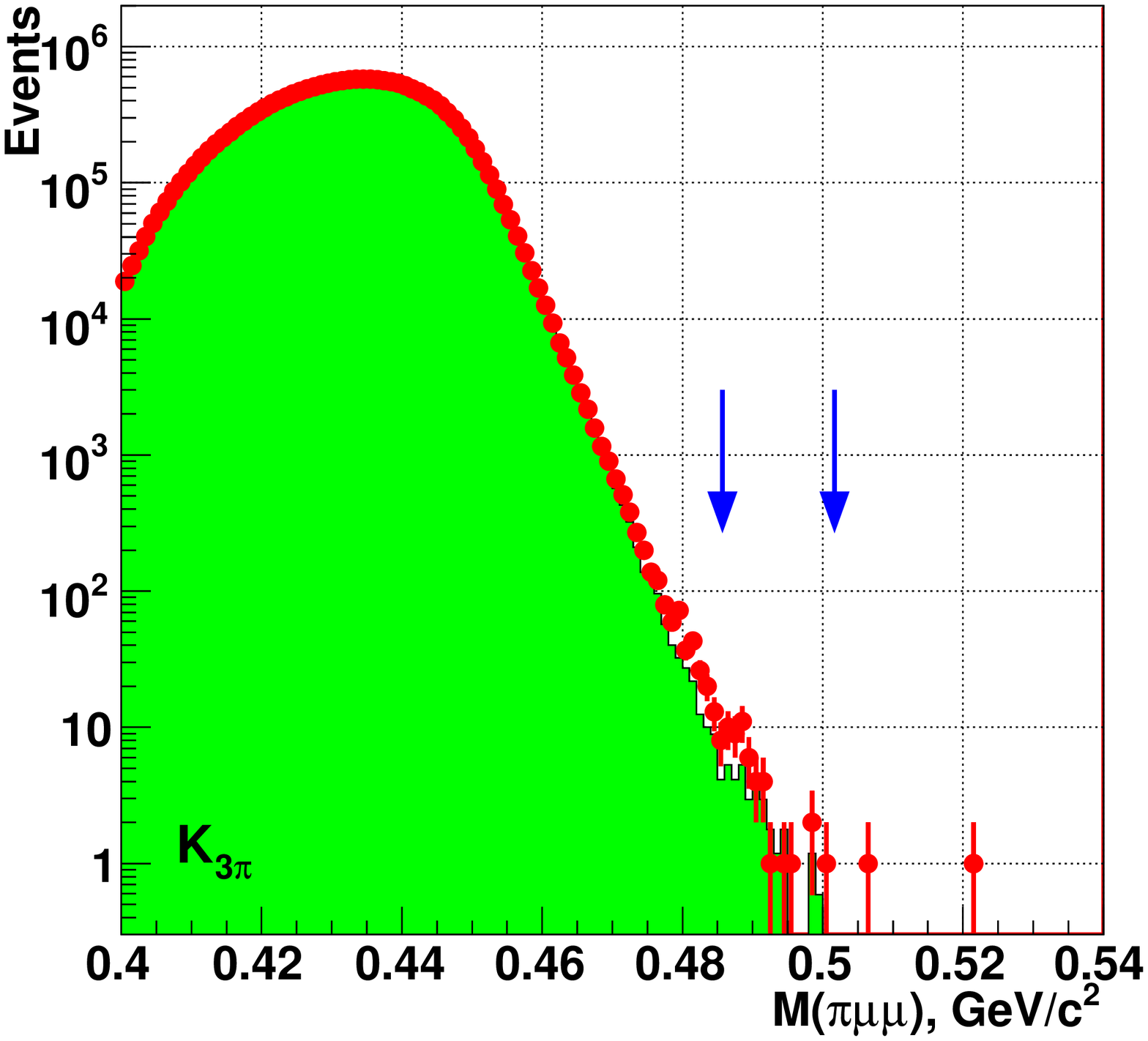}
\caption{Reconstructed $M_{\pi\mu\mu}$ spectra for candidates with different (left) and same sign (right) muons. Dots are the data.}
\label{fig:pmm}
\end{figure}

Figure \ref{fig:pmm} shows the invariant mass spectra of the reconstructed $\pi^\pm\mu^\pm\mu^\mp$ and $\pi^\mp\mu^\pm\mu^\pm$ candidates. The 
$K^\pm\rightarrow\pi^\pm\mu^\pm\mu^\mp$ decay was studied separately \cite{Batley:2011zz}. Simulations showed that $52.6\pm19.8$ events with same sign muons 
were expected in the signal region of the invariant mass spectrum ($|M_{\pi\mu\mu}-M_K|<8$ MeV/c$^2$), mainly due to $K_{3\pi}$ decays: 52 were found
in data. The Feldman-Cousins method \cite{Feldman:1997qc} was applied for the evaluation of the confidence interval, also taking into account the systematic 
uncertainty of the background evaluation. This lead to an upper limit of $BR(K^\pm\rightarrow\pi^\mp\mu^\pm\mu^\pm)<1.1\times10^{-9}$ at 90\% CL, which 
improves the best previous limit by almost of a factor 3.

\section{The ultra-rare kaon decays: future prospects with the NA62 experiment}
\label{sec:pvv}
The goal for the future of the NA62 experiment is the measurement of the branching ratio of the $K^+\rightarrow\pi^+\nu\bar{\nu}$ decay with 10\%
precision. Therefore NA62 aims to collect of the order of 100 $K^+\rightarrow\pi^+\nu\bar{\nu}$ events in about two years of data taking and to 
keep the total systematic uncertainty small. To this purpose, at least $10^{13}$ $K^+$ decays are required, assuming a 10\% signal acceptance and a
$K^+\rightarrow\pi^+\nu\bar{\nu}$ branching ratio of $10^{-10}$. To keep the systematic uncertainty small requires a rejection factor for generic
kaon decays of the order of $10^{12}$, and the possibility to measure efficiencies and background suppression factors directly from data.
In order to match the above required kaon intensity, signal acceptance and background suppression, new detectors must replace the existing NA62 
apparatus.
 
The CERN-SPS extraction line, already used for NA48, can deliver the required intensity, asking for 30\% more SPS protons on target only.
Consequently the NA62 experiment will be housed in the CERN North Area High Intensity Facility (NAHIF) where NA48 was located.
Considerations about signal acceptance drive the choice of a 75 GeV/c charged kaon beam with 1\% momentum bite. The use of a decay-in-flight
technique to identify $K^+$ decay products is the experimental principle of NA62. 

The experimental set-up is close to the one used for NA48: a 100 m beam line to select the appropriate beam, a 80 m evacuated decay
volume and detectors downstream which measure the secondary particles from the kaon decays occurring in the decay volume.

The signature of the signal is one track in the final state matched to one $K^+$ track in the beam. The integrated rate upstream is about 
800 MHz (only 6\% of the beam particles are kaons, the other are $\pi^+$ and protons). The rate seen by the detector downstream is about 10 MHz,
mainly due to $K^+$ decays. Timing and spatial information are needed to match the upstream and downstream track.

Backgrounds come from all the kaon decays with one track left in the final state and from accidental tracks reconstructed downstream matched
by chance to a track upstream. The background suppression profits from the high momentum of the kaon beam. Different techniques have to be employed 
in combination in order to reach the required level of rejection. Schematically they can be divided into: kinematic rejection, precise timing, high 
efficient photon and muon vetoes, precise particle identification systems to distinguish $\pi^+$, $\mu^+$ and positrons. 

The above requirements drove 
the design and the construction of the subdetectors systems. The main subdetectors forming the NA62 layout are: a differential Cerenkov counter
on the beam line to identify the $K^+$ in the beam; a Si-pixel beam tracker; a guard-ring counters surrounding the beam tracker to veto 
catastrophic interactions of particles; a downstream spectrometer made by straw chambers in vacuum; a RICH to distinguish pions and muons; a charged 
hodoscope; a system of photons veto including a series of annular lead glass calorimeters surrounding the decay and detector volume, the NA48 LKr 
calorimeter and a small angle calorimeter to keep the hermetic coverage for photons emitted at zero angle; a muon veto detector.

The design of the experimental apparatus and the R\&D of the new subdetectors was completed in 2010. The experiment is under construction and the first
technical run is foreseen at the end of 2012.

\section{Conclusion}
Kaon decays exhibit good sensitivity to new physics thanks to the high experimental precision achieved. In most of the cases the sensitivity is 
complementary to the one obtained measuring B decays.

The NA62 experiment at CERN provided in 2007 the most precise measurement of the lepton flavour parameter $R_K$: $R_K=(2.487\pm0.013)\times10^{-5}$.
It is consistent with the SM value and can be used to constrain multi-Higgs new physics scenario. NA48/2 improved the upper limit on the branching ratio
of the lepton number violating decay $K^\pm\rightarrow\pi^\mp\mu^\pm\mu^\pm$, which is now $1.1\times10^{-9}$.

The ultra-rare $K\rightarrow\pi\nu\bar{\nu}$ decay is a unique environment where to search for new physics. The NA62 experiment at CERN-SPS 
proposes to follow this road by collecting $O(100)$ events of the $K^+\rightarrow\pi^+\nu\bar{\nu}$ decay. The experiment has been
approved and funded. After three years of successful R\&D program, the NA62 experiment is now under construction.


\begin{thebibliography}{99}

%%
%%  bibliographic items can be constructed using the LaTeX format in SPIRES:
%%    see    http://www.slac.stanford.edu/spires/hep/latex.html
%%  SPIRES will also supply the CITATION line information; please include it.
%%

\bibitem{Deschamps:2009rh}
  O.~Deschamps, S.~Descotes-Genon, S.~Monteil, V.~Niess, S.~T'Jampens and V.~Tisserand,
  Phys.\ Rev.\  D {\bf 82} (2010) 073012.\\
  PHRVA,D82,073012;

\bibitem{Antonelli:2010yf}
  M.~Antonelli {\it et al.},
  Eur.\ Phys.\ J.\  C {\bf 69} (2010) 399.\\
  EPHJA,C69,399;
 
\bibitem{Cirigliano:2007xi}
  V.~Cirigliano and I.~Rosell,
  Phys.\ Rev.\ Lett.\  {\bf 99} (2007) 231801.\\
  PRLTA,99,231801;

\bibitem{Masiero:2005wr}
  A.~Masiero, P.~Paradisi and R.~Petronzio,
  Phys.\ Rev.\  D {\bf 74} (2006) 011701.\\
  PHRVA,D74,011701;  

\bibitem{Masiero:2008cb}
  A.~Masiero, P.~Paradisi and R.~Petronzio,
  JHEP {\bf 0811} (2008) 042.\\
  JHEPA,0811,042;  

\bibitem{Clark:1972nq}
  A.~R.~Clark, B.~Cork, T.~Elioff, L.~T.~Kerth, J.~F.~Mcreynolds, D.~Newton and W.~A.~Wenzel,
  Phys.\ Rev.\ Lett.\  {\bf 29} (1972) 1274.\\
  PRLTA,29,1274;

\bibitem{Heard:1974kk}
  K.~S.~Heard {\it et al.},
  Phys.\ Lett.\  B {\bf 55} (1975) 324.\\
  PHLTA,B55,324;

\bibitem{Heintze:1976qf}
  J.~Heintze {\it et al.},
  Phys.\ Lett.\  B {\bf 60} (1976) 302.\\
  PHLTA,B60,302;

\bibitem{Nakamura:2010zzi}
  K.~Nakamura {\it et al.}  [Particle Data Group],
  J.\ Phys.\ G {\bf 37} (2010) 075021.\\
  JPHGB,G37,075021;

\bibitem{Ambrosino:2009rv}
  F.~Ambrosino {\it et al.}  [KLOE Collaboration],
  Eur.\ Phys.\ J.\  C {\bf 64} (2009) 627
  [Erratum-ibid.\  {\bf 65} (2010) 703]
  [arXiv:0907.3594 [hep-ex]].\\
  EPHJA,C64,627;

\bibitem{Lazzeroni:2011}
  C.~Lazzeroni {\it et al.} [NA62 Collaboration],
  Phys.\ Lett.\  B {\bf 698} (2011) 105.

\bibitem{Zuber:2000vy}
  K.~Zuber,
  Phys.\ Lett.\  B {\bf 479} (2000) 33
  [arXiv:hep-ph/0003160].\\
  PHLTA,B479,33;

\bibitem{Appel:2000tc}
  R.~Appel {\it et al.},
  Phys.\ Rev.\ Lett.\  {\bf 85} (2000) 2877
  [arXiv:hep-ex/0006003].\\
  PRLTA,85,2877;

\bibitem{Batley:2011zz}
  J.~R.~Batley {\it et al.}  [NA48/2 collaboration],
  Phys.\ Lett.\  B {\bf 697} (2011) 107
  [arXiv:1011.4817 [hep-ex]].\\
  PHLTA,B697,107;

\bibitem{Brod:2010hi}
  J.~Brod, M.~Gorbahn and E.~Stamou,
  Phys.\ Rev.\  D {\bf 83} (2011) 034030
  [arXiv:1009.0947 [hep-ph]].\\
  PHRVA,D83,034030;

\bibitem{Isidori:2006qy}
  G.~Isidori, F.~Mescia, P.~Paradisi, C.~Smith and S.~Trine,
  JHEP {\bf 0608} (2006) 064
  [arXiv:hep-ph/0604074].\\
  JHEPA,0608,064;

\bibitem{Blanke:2009am}
  M.~Blanke, A.~J.~Buras, B.~Duling, S.~Recksiegel and C.~Tarantino,
  Acta Phys.\ Polon.\  B {\bf 41} (2010) 657
  [arXiv:0906.5454 [hep-ph]].\\
  APPOA,B41,657;

\bibitem{Artamonov:2009sz}
  A.~V.~Artamonov {\it et al.}  [BNL-E949 Collaboration],
  Phys.\ Rev.\  D {\bf 79} (2009) 092004
  [arXiv:0903.0030 [hep-ex]].\\
  PHRVA,D79,092004;

\bibitem{Anelli:2005ju}
  G.~Anelli {\it et al.},
  CERN-SPSC-P-326;

\bibitem{Batley:2007yfa}
  J.~R.~Batley {\it et al.}  [NA48/2 Collaboration],
  Eur.\ Phys.\ J.\  C {\bf 52} (2007) 875
  [arXiv:0707.0697 [hep-ex]].\\
  EPHJA,C52,875;

\bibitem{Fanti:2007vi}
  V.~Fanti {\it et al.}  [NA48 Collaboration],
  Nucl.\ Instrum.\ Meth.\  A {\bf 574} (2007) 433.\\
  NUIMA,A574,433;

\bibitem{Agostinelli:2002hh}
  S.~Agostinelli {\it et al.}  [GEANT4 Collaboration],
  Nucl.\ Instrum.\ Meth.\  A {\bf 506} (2003) 250.\\
  NUIMA,A506,250;

\bibitem{Bijnens:1992en}
  J.~Bijnens, G.~Ecker and J.~Gasser,
  Nucl.\ Phys.\  B {\bf 396} (1993) 81
  [arXiv:hep-ph/9209261].\\
  NUPHA,B396,81;

\bibitem{Weinberg:1965nx}
  S.~Weinberg,
  Phys.\ Rev.\  {\bf 140} (1965) B516.\\
  PHRVA,140,B516;

\bibitem{Gatti:2005kw}
  C.~Gatti,
  Eur.\ Phys.\ J.\  C {\bf 45} (2006) 417
  [arXiv:hep-ph/0507280].\\
  EPHJA,C45,417;

\bibitem{Feldman:1997qc}
  G.~J.~Feldman and R.~D.~Cousins,
  Phys.\ Rev.\  D {\bf 57} (1998) 3873
  [arXiv:physics/9711021].\\
  PHRVA,D57,3873;

\end{thebibliography}
\end{document}